\documentclass[conference]{IEEEtran}
\IEEEoverridecommandlockouts

\usepackage{xcolor}
\usepackage[normalem]{ulem}

\newcommand{\MOD}[1]{\textcolor{black}{#1}}

\usepackage{hyperref}
\usepackage[
  backend=biber,
  style=ieee,
  maxbibnames=99,
  minbibnames=99
]{biblatex}

\usepackage[ruled]{algorithm2e}
\usepackage{float}
\usepackage{microtype}
\usepackage{enumitem}

\usepackage{amsmath,amssymb,amsfonts, bm}
\usepackage{algorithmic}
\usepackage{graphicx}
\usepackage{textcomp}
\usepackage{xcolor}
\def\BibTeX{{\rm B\kern-.05em{\sc i\kern-.025em b}\kern-.08em
    T\kern-.1667em\lower.7ex\hbox{E}\kern-.125emX}}
    
\DeclareMathOperator*{\argmin}{argmin}

\DeclareMathOperator*{\argmax}{argmax}

\DeclareMathOperator{\Id}{Id}
\DeclareMathOperator{\prox}{prox}
    
\begin{document}

\title{Multiresolution Adaptive Block-Coordinate Forward--Backward for Image Reconstruction\\
}

\author{\IEEEauthorblockN{Edgar Desainte-Maréville\textsuperscript{1}, Marion Foare\textsuperscript{1,3}, Paulo Gonçalves\textsuperscript{1}, Nelly Pustelnik\textsuperscript{2}, Elisa Riccietti\textsuperscript{1}}
\IEEEauthorblockA{\textit{\textsuperscript{1}ENS de Lyon, CNRS, Inria, LIP, UMR 5668, Université Claude Bernard Lyon 1, Lyon, France} \\
\textit{\textsuperscript{2}CNRS,  ENS de Lyon,  Laboratoire de Physique, UMR 5672,  Lyon, France}\\
\textit{\textsuperscript{3}CPE Lyon, Villeurbanne, France} \\
\{edgar.desainte-mareville, 
marion.foare, 
paulo.goncalves, 
nelly.pustelnik,
elisa.riccietti\}@ens-lyon.fr}
}

\maketitle

\begin{abstract}

Classical first-order optimization methods for imaging inverse problems scale poorly with image resolution. Wavelet-based multilevel strategies can accelerate convergence under strong blur, but their fixed coarse-to-fine schedules lose effectiveness in moderate-blur or noise-dominated regimes.

In this work, we propose an adaptive multiresolution block-coordinate Forward--Backward algorithm for image restoration.
Multiresolution block selection is driven by the local magnitude of the proximal update via a stochastic non-smooth Gauss--Southwell rule applied to the wavelet decomposition of the image.
This adaptive selection strategy dynamically balances updates across scales, emphasizing coarse or fine blocks according to the degradation regime.
As a result, the proposed method automatically adapts to varying blur and noise levels without relying on a predefined hierarchical update scheme.
\end{abstract}

\begin{IEEEkeywords}
Inverse problems, image restoration, multiscale optimization, block-coordinate Forward--Backward, Gauss--Southwell rule
\end{IEEEkeywords}

\section{Introduction}

Image restoration aims at recovering a ground truth image from a degraded observation.
Such problems are commonly formulated within a variational framework, leading to large-scale convex optimization problems of the form
\begin{equation}
\label{eq:main}
\min_{x \in \mathbb{R}^n} \; f(x) + h(x),
\end{equation}
where $f$ is a smooth data-fidelity term and $h$ is a possibly non-smooth regularization term.
First-order methods based on proximal splitting, such as the Forward--Backward (FB) scheme \cite{combettes2005signal, combettes2011proximal} and its accelerated variants \cite{beck2009fast}, are widely used due to their simplicity and robustness.
These approaches remain central in model-based neural network methods, such as Plug-and-Play (PnP) frameworks, where a trained denoiser replaces the proximal operator.
Among these, Forward--Backward PnP (FB-PnP) \cite{pnp, hurault2022gradient,pesquet2021learning} stands out as one of the most widely used hybridizations of neural networks and variational techniques.

Despite their strong theoretical guarantees, these methods often exhibit slow convergence, which becomes limiting when applied to high-resolution images.
Improving the scalability of optimization algorithms with respect to image resolution is therefore a central challenge in imaging inverse problems.

A standard strategy to address large-scale optimization problems consists in decomposing the variables into smaller subsets, referred to as blocks, and performing block-wise updates.
Block-coordinate descent (BCD) methods proceed by iteratively updating one or several blocks of variables while keeping the remaining components fixed \cite{tseng2001convergence}.

When the objective function takes the form \eqref{eq:main} the resulting family of methods is referred to as block-coordinate Forward--Backward (BC-FB), and each method is determined by the block selection strategy.
A simple choice consists in updating all the blocks at each iteration, which recovers the standard Forward--Backward algorithm.
Alternatively, stochastic BC-FB methods rely on random block activations, typically drawn independently at each iteration according to a fixed probability distribution, and enjoy theoretical convergence guarantees \cite{nesterov2012efficiency, salzo2022parallel}.
However, such uniform sampling strategies do not exploit the structure of the problem and may lead to minor improvements.
More recently, the classical deterministic Gauss--Southwell rule \cite{southwell}, which selects at each iteration the block associated with the largest partial gradient norm, has been extended to a non-smooth context by selecting the block associated with the largest proximal update \cite{nutini2015coordinate}.

While block-coordinate strategies are conceptually appealing, a direct extension of these updates to imaging inverse problems would lead to patch-induced block decompositions, which can complicate the formulation of the subproblems, especially in the presence of global regularization terms, due to boundary effects and interactions between blocks \cite{pascal_block-coordinate_2018}.

A more natural way to build a block decomposition for images is to exploit their intrinsic multiscale structure by leveraging  wavelet-based decompositions \cite{chouzenoux2016block}. Specifically,
recent works in multilevel optimization \cite{lauga2024iml, briceno2025flexible} proposed a Multilevel Forward--Backward (MLFB) method that chooses blocks corresponding to approximation and detail coefficients across scales, and performs BC-FB updates according to fixed coarse-to-fine schedules, where coarse scales correspond to wavelet approximation coefficients.
This approach can be interpreted as a BC-FB method with a predetermined hierarchical update order, and has been shown to accelerate convergence, especially in regimes dominated by strong blur and low noise.

However, when the blur is moderate or when the noise becomes dominant, these fixed coarse-to-fine update rules tend to lose effectiveness.
This observation suggests that no single fixed ordering is optimal across different degradation regimes, and motivates the design of adaptive block-selection strategies capable of dynamically identifying the components that will contribute the most to improving the reconstruction during the optimization process.

\noindent \textbf{Contribution} -- In this work, we propose a Multiresolution Adaptive Gauss-Southwell Inspired Coordinate Forward--Backward (\texttt{MAGIC-FB}) algorithm for image restoration.
Block selection is inspired by a proximal Gauss-Southwell rule applied across resolution levels rather than spatial patches, which, to the best of our knowledge, has never been explored in imaging.
Moreover, to allow for parallel block updates, we propose a stochastic variant of the Gauss-Southwell selection rule, with probability distributions driven by the local magnitude of the proximal update.
Particular attention is paid to the computation of partial gradients, which leverages the linearity of the problem to enable cheap updates.

This adaptive strategy dynamically balances updates across scales, selecting coarse or fine components depending on the degradation type.
As a result, the proposed method automatically adapts to varying blur and noise levels without relying on a predefined classical multilevel update schedule, thus making the proposed method a good choice in any degradation scenario.

We demonstrate indeed through numerical experiments that, across a wide range of blur and noise regimes, the proposed strategy consistently matches or outperforms the best methods, whose performance are on the contrary heavily affected by the degradation type. \MOD{Convergence guarantees for this heuristic scheme are left for future work.}
\\

\noindent \textbf{Notations} --
The Euclidean norm is denoted by $\|\cdot\|$.   For a convex, proper, and lower semicontinuous function $g$, the proximal operator of $g$ is defined as
$
\prox_{g}(z) := \argmin_{u \in \mathbb R^n} \left\{ g(u) + \frac{1}{2}\|u - z\|^2 \right\}.
$
For a block-structured vector $w$, $w_i$ denotes its $i$-th block, while superscripts denote iterations indices.
For a function $f : \mathbb R^n \to \mathbb{R}$, $\nabla f$ denotes its gradient when it exists and $\nabla_{w_i} f(w)$ denotes the partial gradient with respect to the block $w_i$.

\vspace{-0.1cm}
\section{Wavelet-based block-coordinate Forward--Backward}

\subsection{Wavelet-based imaging inverse problem}

We consider an imaging inverse problem where an unknown image $\overline x \in \mathbb R^n$ is observed through a linear degradation operator $A \in \mathbb R^{m \times n}$ and corrupted by additive Gaussian noise:
\begin{equation}
    y = A\overline x + \eta,
\end{equation}
where  $\eta \sim \mathcal{N}(0, \sigma^2 \Id)$.
In this work, we focus on variational approaches formulated as in \eqref{eq:main}, where the penalization is expressed in the wavelet domain.
The wavelet-based formulation was extremely popular a few decades ago, and receives nowadays a renewed interest for its capability of improving the reconstruction performance when combined with neural networks, as recently highlighted in \cite{kadkhodaie2023learning, laurent2025multilevel}, where the multiscale structure plays a crucial role in achieving high-quality restoration.

We thus focus on a minimization problem expressed as:
\begin{equation}
    \label{optim_wavelet}
    \widehat w \in \underset{w \in \mathbb R^n}{\mathrm{Argmin}}\; \varphi(w)
    := \underbrace{\tfrac{1}{2}\| AW^\top w - y\|^2}_{=:{f}(w)}
    + g(w),
\end{equation}
where $W \in \mathbb{R}^{n \times n}$ is an orthonormal wavelet transform with $J$ decomposition levels and the reconstructed image is  $\widehat x = W^\top \widehat w$.
The function $w \mapsto \tfrac{1}{2}\|AW^\top w - y\|^2$ is $L$-smooth with
$L = \|AW^\top\|^2 = \|A\|^2$.

The regularization term $g : \mathbb R^n \to \mathbb R$ is assumed to be convex, proper,
and lower semicontinuous, and is typically non-smooth.
It is well established in the literature \cite{chaari2009solving} that, for an orthonormal wavelet transform, problem~\eqref{optim_wavelet} is  equivalent to \eqref{eq:main} with $h = g(W\cdot)$.

\subsection{Wavelet representation and separable regularization}

Wavelet transforms provide a multiresolution representation of an image, which is decomposed into low-frequency and high-frequency components.

To avoid distinguishing between approximation and detail coefficients and to emphasize the block structure induced by the wavelet transform, we introduce the unified notation
\begin{align*} Wx &= [a_J^\top, d_J^\top, \ldots, d_1^\top]^\top =[w_0^\top, w_1^\top, \ldots, w_J^\top]^\top=w
\end{align*}
where $a_J = w_0 \in \mathbb R^{n_0}$ denotes the approximation coefficients at the coarsest scale, and $d_j = w_{J - j + 1} \in \mathbb R^{n_{J-j+1}}$ contains the detail coefficients at scale $j$, formed by concatenating the horizontal, vertical, and diagonal sub-bands.

We assume that the regularization term is separable across wavelet blocks, \textit{i.e.}, that its contribution on each block is given by proper, convex, lower semicontinuous functions $g_i\colon \mathbb{R}^{n_i} \to (-\infty,+\infty]$ for all $i\in \{0,\ldots, J\}$ such that
\begin{equation}
\label{eq:separable}
(\forall w \in \mathbb R^n) \qquad  g(w) = \sum_{i=0}^J g_i(w_i).
\end{equation}

\subsection{General algorithm and state-of-the-art update rules}

The BC-FB method designed to solve \eqref{optim_wavelet} when $g$ is defined by \eqref{eq:separable} exploiting a multiresolution architecture \cite{briceno2025flexible} is detailed in Algorithm \ref{BCDalg}. 

\SetAlgoNoEnd
\begin{algorithm}[h]
\caption{BC-FB for \eqref{optim_wavelet}}
\label{BCDalg}
\KwIn{$w^{[0]}\in \mathrm{dom}\,\varphi$, step $\gamma\in(0,\tfrac{2}{L})$,
$\varepsilon^{[k]}\in\{0,1\}^{J+1}$}
\For{$k=0,1,\dots$}{
  \For{$i=0,\dots,J$}{
    \eIf{$\varepsilon_i^{[k]} = 1$}{
      $w_i^{[k+1]}=\prox_{\gamma g_i}\!\big(w_i^{[k]}-\gamma\nabla_i f(w^{[k]})\big)$\;
    }{
      $w_i^{[k+1]}=w_i^{[k]}$\;
    }
  }
}
\end{algorithm}

The sequence $(\varepsilon^{[k]})_{k \in \mathbb N}$ encodes the update rule: for each $k$, block $i$ is updated if $\varepsilon_i^{[k]} = 1$, and left unchanged otherwise.
The update rules of the state-of-the-art are the following:
\begin{itemize}
    \item \textit{Forward-Backward \cite{combettes2005signal}:}
    All blocks are updated at each iteration, \textit{i.e.}, $\varepsilon^{[k]}_i = 1$, $\forall k \in \mathbb N$ and $\forall 0\leq i\leq J$.

    \item \textit{Uniform stochastic sampling:}
    At each iteration $k$ and for each block $i$, the update variable
    $\varepsilon_i^{[k]}$ is drawn independently according to a Bernoulli
    distribution with a fixed parameter $p$, \textit{i.e.}, $\varepsilon_i^{[k]} \sim \mathcal{B}(p).$

    \item \textit{Multilevel Forward-Backward \cite{lauga2024iml}:}
    The MLFB strategy follows a deterministic coarse-to-fine update pattern with
        $\varepsilon^{[k]}_i = 1$ $\forall i < k \mod (J+1)$.
\end{itemize}

The next section describes our proposed adaptive block selection strategy.

\section{MAGIC-FB}

\subsection{Proposed multiresolution Gauss--Southwell update }

We now introduce MAGIC-FB, an adaptive multiresolution block-coordinate FB algorithm based on a new stochastic Gauss--Southwell selection rule.
The method fits within the general framework in Algorithm~\ref{BCDalg} and corresponds to a specific choice of the block-update sequence
$(\varepsilon^{[k]})_{k\in\mathbb N}$.

Given the current iterate $w^{[k]} = (w_0^{[k]},\ldots,w_J^{[k]})$ and a stepsize $\gamma > 0$, we define
for each block $0 \le i \le J$ the block-wise proximal update
\vspace{-0.3cm}

$$
\Delta_i^{[k]}
:= w_i^{[k]} - \prox_{\gamma g_i}\!\left(w_i^{[k]} - \gamma \nabla_{w_i} f\left(w^{[k]}\right)\right),
$$
\vspace{-0.3cm}

\noindent which quantifies the variation induced by a proximal gradient step restricted to block $i$.
At iteration $k$, the deterministic Gauss-Southwell rule would select the block index $i^{[k]} \in \argmax_{0 \le i \le J} \|\Delta_i^{[k]}\|$.


To leverage the flexibility of randomness and to allow for parallel updates, we introduce a stochastic variant by defining activation probabilities
$p_i^{[k]} = \frac{\|\Delta_i^{[k]}\|}{\|\Delta^{[k]}\|}, \qquad
\|\Delta^{[k]}\| := \Big(\sum_{i=0}^J \|\Delta_i^{[k]}\|^2\Big)^{1/2},
$
which satisfy $0 \le p_i^{[k]} \le 1$.
Each block is then independently activated according to
$\varepsilon_i^{[k]} \sim \mathcal B(p_i^{[k]})$,
with higher probability thus assigned to blocks associated with larger proximal updates. Another advantage of this update rule is its flexibility: it does not require fixing in advance the number of blocks updated at each iteration, allowing this number to change from one iteration to the next.

\subsection{Computational efficiency of MAGIC-FB}

At first glance, the greedy Gauss-Southwell rule can seem too expensive to effectively reduce the convergence time, since it requires to compute  $\Delta_i^{[k]}$ for all the indices $i$. We detail in this subsection how, in the context of linear inverse problems, the structure of the objective function allows for fast computation of these quantities.
Let $w \in \mathbb R^n$ be an image expressed in the wavelet domain. We denote by $S_i$ the wavelet subspace associated with the block $i$ of size $d_i$, and $\Pi_i : \mathbb R^n \to S_i$ the wavelet projection operator.
For any image  $x \in \mathbb R^n$,  $w_i = \Pi_i x$.
We have
{\small{
\begin{align*}
    \nabla f(w) &= (AW^\top)^\top (AW^\top w - y) \\
    &= W A^\top A \big(\sum_{j = 0}^J \Pi_j^\top w_j\big) - W A^\top y \\
    &= \sum_{j=0}^J W A^\top A \Pi_j^\top w_j - W A^\top y.
\end{align*}}}
The partial gradient with respect to the block $0\leq i \leq J$ can therefore be written as
$$ \nabla_{w_i} f(w) = \sum_{j=0}^J \Pi_i A^\top A \Pi_j^\top w_j - \Pi_i A^\top y.$$
The matrices $(\Pi_i A^\top A \Pi_j^\top)_{0 \leq i,j \leq J}$ and the constant terms $(\Pi_i A^\top y)_{0 \leq i \leq J}$ can be pre-computed and stored.

When evaluating $\Delta_i^{[k]}$ at iteration $k$, all the terms $(\Pi_i A^\top A \Pi_j^\top w_j^{[k]})_{0 \le i,j \le J}$ can be  stored.
If block $i^{[k]}$ is chosen and updated, in order to update the value of the gradient for the next iteration, it is sufficient to recompute the matrix-vector products of the terms that depend on block $i^{[k]}$
$$ (\Pi_i A^\top A \Pi_{i^{[k]}}^\top w^{[k+1]}_{i^{[k]}})_{0 \leq i \leq J}, $$ as all the other partial products $(\Pi_i A^\top A \Pi_j^\top w_j^{[k]})_{0 \le i,j \le J}$ remain unchanged, since $w_j^{[k]}=w_j^{[k+1]}$ for $j\neq i^{[k]}$.

This allows the partial gradients $(\nabla_{w_i} f(w))_{0 \le i \le J}$ to be evaluated efficiently from the current block coefficients $(w_i)_{0 \le i \le J}$ without recomputing global matrix--vector products, thus making
the block-selection step computationally inexpensive.

As a result, the proposed greedy block-selection strategy does not increase the complexity compared to standard block-coordinate proximal gradient methods.
Instead, it reallocates computational effort by selecting the most relevant block before performing the costly matrix--vector products associated with the update, thereby improving overall efficiency.

\vspace{-0.1cm}
\section{Experiments}
\vspace{-0.1cm}

In this section, we evaluate the performance of the proposed algorithm, which is assessed using a performance profile  \cite{dolan_benchmarking_2002}  shown on Figure~\ref{fig:profiles}, complemented by representative extreme cases for qualitative analysis on Figure~\ref{fig:curves}. All experiments were conducted on one of the Blaise Pascal Center's machines \cite{sidus}, equipped with an NVIDIA RTX 3090 Ti GPU, an Intel Xeon Gold 6544Y CPU, and 125 GB of RAM. Code is available at \url{https://github.com/EdgarDesainteMareville/MAGIC-FB}.

\vspace{-0.2cm}
\subsection{Experimental setup}
\vspace{-0.2cm}

\paragraph{Test dataset}
Experiments are conducted on $1024 \times 1024$ cropped regions extracted from $100$ images of the DIV2K dataset \cite{Agustsson_2017_CVPR_Workshops}. For each test instance, the blur and noise levels are randomly drawn to span a broad range of degradation conditions. 
\paragraph{Degradation model}
The forward operator $A$ corresponds to a spatially invariant Gaussian blur with variance $\sigma_\mathrm{blur}$ implemented as a separable convolution with circular boundary conditions. White Gaussian noise with variance $\sigma_\mathrm{noise}$ is added to the observations.
Blur levels are uniformly sampled with $\sigma_{\mathrm{blur}} \in [1,15]$ and noise levels are log-uniformly sampled with $\sigma_{\mathrm{noise}} \in [10^{-3},10^{-1}]$.

\MOD{Additional experiments on larger images and with deeper wavelet decompositions yielded consistent results; they are omitted here due to space constraints.}

\paragraph{Multiresolution representation and regularization}
We use an orthonormal Daubechies-8 wavelet transform with $J = 5$ levels to decompose the images\MOD{, yielding a $32 \times 32$ approximation subband, coarse enough to expose multiscale structure while retaining the global content of the image.}
We use the classical sparsity promoting $\ell_1$ norm of the wavelet coefficients as a regularization term, balanced by a regularization coefficient $\lambda > 0$: for all $i\in\{1,\ldots, J\}$, $g_i = \lambda \|\cdot\|_1$ and $g_0 \equiv 0$.
This choice naturally satisfies \eqref{eq:separable}.
Note that the approximation coefficients $w_0$ are not penalized, as they are typically not sparse.
The regularization parameter is shared across all scales and is selected by a grid search \MOD{over a logarithmic grid of 20 values ranging from $10^{-5}$ to $1$} to maximize the peak signal-to-noise ratio (PSNR) with respect to the ground truth.

\paragraph{Algorithms and parameters}
We compare the following optimization schemes:
\begin{description}[font=\normalfont]
    \item[\texttt{FB}:] classical forward-backward splitting \cite{combettes2005signal}.
    \item[\texttt{Stoc-BC-FB}:] uniform sampling stochastic multiresolution BC-FB method, where each block has a $50\%$ chance of being updated at each iteration.
    \item[\texttt{MLFB}:] Multilevel Forward-Backward -- fixed hierarchical coarse-to-fine multiresolution BC-FB from \cite{briceno2025flexible}.
    \item[\texttt{MAGIC-FB}:] the proposed stochastic multiresolution adaptive proximal Gauss--Southwell BC-FB method.
\end{description}
All the methods are initialized with the observed image $w^{[0]} = Wy$ and use the same stepsize $\gamma = \tfrac{1.9}{\|A\|^2},$
where $\|A\|$ is estimated numerically.
All methods are run for 200 iterations, which are sufficient for the methods to converge in all the considered settings.

\vspace{-.15cm}
\subsection{Numerical results}
\vspace{-.1cm}
\begin{figure*}[t]
    \centering
    \setlength{\tabcolsep}{5pt}
    \begin{tabular}{cc}
       \includegraphics[width=0.98\textwidth]{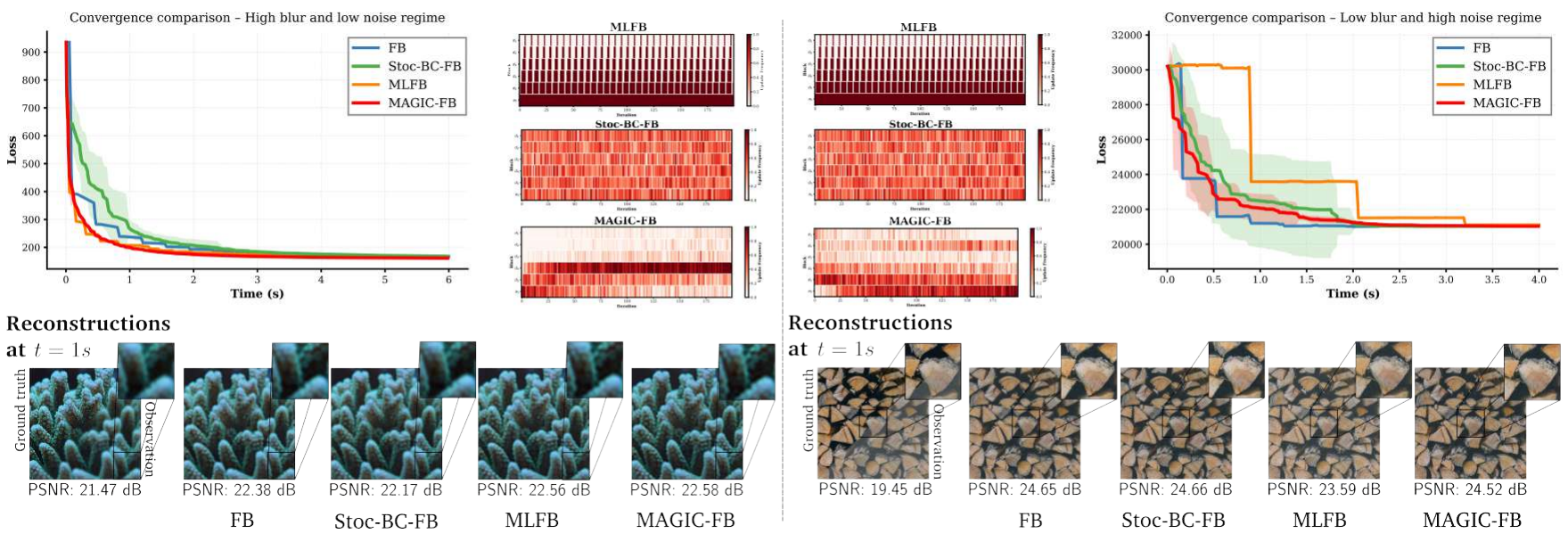}%
    \end{tabular} 
    \vspace{-.4cm}
    \caption{
    Convergence behavior and block-selection patterns for the considered methods on an image deblurring problem.
    Left: $\sigma_\mathrm{blur} = 7$ and $\sigma_\mathrm{noise} = 0.01$.
    Right: $\sigma_\mathrm{blur} = 1$ and $\sigma_\mathrm{noise} = 0.1$.
    Top left and right panels: objective function value as a function of time.
    Centered top panels: block activation patterns across wavelet scales.
    Bottom panels: image reconstructions at $t = 1\,\mathrm{s}$ for each method.
    }
    \label{fig:curves}
    \vspace{-.6cm}
\end{figure*}
Figure~\ref{fig:curves} reports the convergence behavior of the different optimization schemes for two extreme degradation regimes, in terms of the objective function value as a function of wall-clock time. 
The time required to precompute the matrix products $(\Pi_i A^\top A \Pi_j)_{i,j}$ is excluded from the reported runtimes, as these quantities are shared by all methods. 
For the stochastic methods, results are reported as mean and standard deviation over 10 independent runs.

In the strong blur / low noise regime where $\sigma_\mathrm{blur} = 7$ and $\sigma_\mathrm{noise} = 0.01$, \texttt{MLFB} outperforms \texttt{FB} and \texttt{Stoc-FB}, highlighting the relevance of coarse-scale updates when the degradation is a low-pass filter, as the gain associated with the fitting term outweighs the decrease in the objective function.
In this setting, \texttt{MAGIC-FB} closely matches the behavior of \texttt{MLFB}, while \texttt{Stoc-FB} is ineffective.

In the low blur / high noise regime, where $\sigma_\mathrm{blur} = 1$ and $\sigma_\mathrm{noise} = 0.1$, \texttt{FB} is the most effective method. While \texttt{MLFB} does not perform well, the stochastic methods \texttt{Stoc-FB} and \texttt{MAGIC-FB} achieve fast convergence, showing the importance of parallel updates and exploration allowed by randomness.

The proposed \texttt{MAGIC-FB} method shows to be able
to adapt to all degradation regimes, in contrast to its competitors whose performance strongly vary with the scenario.

The center top panels of Figure~\ref{fig:curves} show the empirical block activation frequencies for the multiresolution methods (excluding \texttt{FB}) with respect to the iterations. For the stochastic update rules, activation frequencies are averaged across runs. These heatmaps provide insight into how block selection evolves under different degradation regimes.

In the strong blur / low noise regime, \texttt{MAGIC-FB} 
essentially activates coarse-scale blocks, evidencing that, at the beginning, there is little to be  gained by  updating fine-scale blocks.
Then, \texttt{MAGIC-FB} gradually shifts towards the fine scale updates, naturally reproducing the \texttt{MLFB} updating rule, and achieving comparable performance.

Conversely, in the low blur / high noise regime, block activations are more evenly distributed across scales, suggesting that restoring simultaneously blocks at all scales contributes to improving the objective function.

Intermediate reconstructions at a fixed time budget of $t = 1\,\mathrm{s}$, together with their corresponding PSNR values, are displayed to provide a qualitative comparison of reconstruction quality across methods.

\begin{figure}
    \centering
    \includegraphics[width=0.8\columnwidth, trim= .2cm .2cm .2cm .2cm, clip]{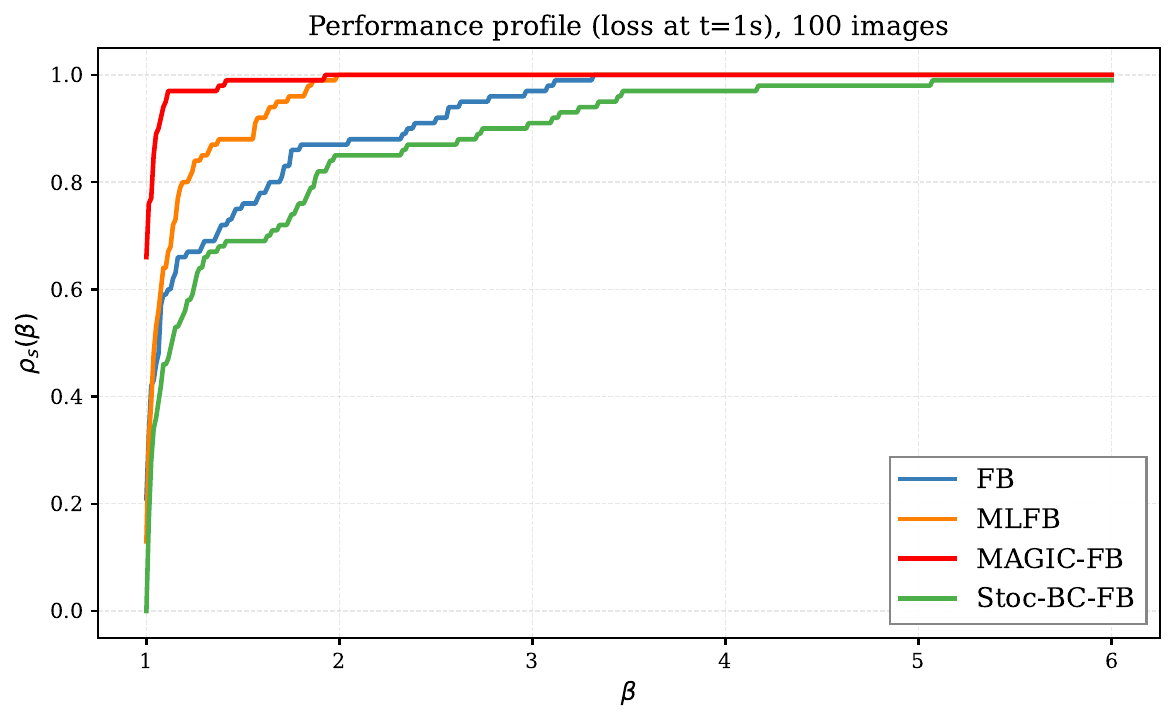}
    \vspace{-.4cm}
    \caption{Performance profiles of the considered methods across the 100 test instances. The $x$-axis represents the factor $\beta$  by which a method's performance is compared to the best method for each problem, while the $y$-axis shows the proportion of problems for which the method is within this factor of the best performance (the closer to the upper left corner, the best).\vspace{-0.6cm}}
    \label{fig:profiles}
\end{figure}

The performance profile shown on Figure~\ref{fig:profiles} provides an evaluation of the methods efficiency across the test set of 100 DIV2K images, illustrating the proportion of problems for which each method is within a certain factor of the best performance, measured in terms of loss function value obtained after 1 second of execution. The proposed \texttt{MAGIC-FB} strategy attains the lowest objective value within the prescribed time budget on over 65\% of the test instances, while maintaining strong robustness across performance ratios compared to the other methods.

Overall, these results show that the proposed strategy dynamically reallocates computational effort across scales and recovers, in each regime, a block-selection pattern consistent with the structure of the underlying inverse problem.

\vspace{-0.1cm}
\section{Conclusion}
\vspace{-0.1cm}

In this work, we introduced \texttt{MAGIC-FB}, an adaptive multiresolution block-coordinate Forward--Backward method for image restoration problems. By leveraging a wavelet-based block decomposition and an adaptive block-selection driven by update magnitudes via a stochastic proximal Gauss--Southwell rule, the proposed algorithm automatically balances updates across scales. This allows for an automatic adaptation of the algorithm to different degradation regimes, ranging from strongly blurred to noise-dominated images, without relying on a fixed  schedule.

Experimental results on image deblurring problems highlight the limitations of fixed update rules in multiresolution strategies when blur or noise conditions vary, and demonstrate that the proposed stochastic approach matches the best methods in the two extreme regimes. In particular, the adaptive mechanism recovers coarse-to-fine behavior in strongly blurred settings, while progressively adding fine-scale updates as noise becomes dominant.

\MOD{Future work will investigate more powerful regularizers such as those used in PnP, aiming for strategies that are theoretically grounded, computationally efficient, with state of the art reconstructions. Such an extension requires a theoretical analysis under the more challenging setting of weak convexity.}

\vspace{-0.15cm}
\section*{Acknowledgment}
\vspace{-0.15cm}

This work was funded by the ANR-24-CE23-7039 MEPHISTO project, the Simone and Cino Del Duca foundation and the PEPR IA project.
We thank the Blaise Pascal Center for its computational support, using the SIDUS \cite{sidus} solution.

\vspace{-0.2cm}

\printbibliography


\end{document}